\documentclass{JINST}
\let\ifpdf\relax

\title{Light propagation and fluorescence quantum yields in liquid scintillators}

\author{C.~Buck\thanks{Corresponding author.}, B.~Gramlich, S.~Wagner\footnote{now at Centro Brasileiro de Pesquisas Fisicas, Rio de Janeiro, RJ, 22290-180, Brazil}\\
\llap{}Max-Planck-Institut f\"ur Kernphysik,\\
  Saupfercheckweg 1, 69117 Heidelberg, Germany\\

E-mail: \email{christian.buck@mpi-hd.mpg.de}}

\abstract{For the simulation of the scintillation and Cherenkov light propagation in large liquid scintillator detectors a detailed knowledge about the absorption and emission spectra of the scintillator molecules is mandatory. Furthermore reemission probabilities and quantum yields of the scintillator components influence the light propagation inside the liquid. Absorption and emission properties are presented for liquid scintillators using 2,5-Diphenyloxazole (PPO) and 4-bis-(2-Methylstyryl)benzene (bis-MSB) as primary and secondary wavelength shifter. New measurements of the quantum yields for various aromatic molecules are shown.}


\begin{document}

\section{Introduction}
In many particle physics experiments using organic liquid scintillators, precise knowledge on the parameters affecting the propagation of scintillation light inside the detector is required. This knowledge is needed to estimate the detector response and to simulate detector signals under various conditions. In particular, for large liquid scintillator (LS) detectors in the ton-scale basic properties of the LS molecules as the absorption and emission spectra or the fluorescence quantum yields need to be known. Such detectors are for example used in neutrino physics, since the cross sections for neutrino detection reactions are tiny and therefore large target masses are needed. The understanding of light propagation in combination with calibration data is paving the way for the determination of the energy deposited inside the detector by ionizing particles. 

In chapter 2 we describe the light absorption and emission properties of some substances typically used in LS neutrino experiments. In databases these spectra are normally given assuming low concentrations and an inert medium around the fluorescent molecules. Both conditions are not fulfilled for standard liquid scintillators which can be affected by solvent and concentration effects. Therefore the spectra as measured in the corresponding solvent and at a given concentration might be more relevant. In addition, the spectrum can change as the light propagates through the medium.  

In chapter 3 we present results of our new measurements on the fluorescence quantum yields for several molecules used in various LS neutrino detectors. It is hard to control the systematic uncertainties in such measurements, which might be the reason why the literature values of the absolute numbers are spread over a wide range with significant deviations from each other. The agreement between quantum yield ratios of molecules measured in our setup compared to the corresponding ratios in other publications is found to be more robust.  

\section{Scintillator absorption and emission}

Ionizing particles traversing liquid scintillators mainly excite the solvent molecules of the liquid. Deexcitation can occur essentially via the emission of fluorescence light or non-radiative energy transfer to another molecule. Since the aromatic solvent molecules are not transparent for their own flourescence light, in particular in large scale LS detectors, primary and sometimes secondary wavelength shifters are added to the liquid. Most modern experiments use PPO (2,5-Diphenyloxazole) as primary and bis-MSB (4-bis-(2-Methylstyryl)benzene) as secondary wavelength shifter. 

\begin{figure}[tb]
\begin{center}
\includegraphics[scale=0.50]{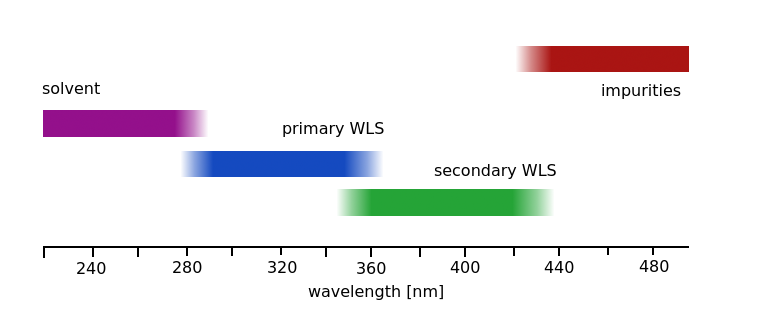}
\end{center}
\caption[]{In an organic liquid scintillator system with a primary and secondary wavelength shifter the absorption bands of the individual components should be reasonably separated. The bars in the plot show for each component the wavelength region for which its own absorbance is dominating in the scintillator mixture.}
\label{Fig1}
\end{figure}

For the modelling of light propagation inside the LS it is important to know on which molecule the photon is absorbed. UV light in wavelength regions below 280 nm is mainly absorbed by the solvent molecules. Benzene derivatives typically have absorption bands around 260~nm and reemit light around 300~nm~\cite{Ber}. Ideally the region of solvent emission matches the absorption peak of the primary wavelength shifter which dominates the absorption in the mixture from around $280-350$~nm. The scintillation light seen by the photomultipliers in a large scale LS detector is above 350~nm. In this region the secondary wavelength shifter is typically dominating the absorption. Therefore, it is very important to have a high quantum yield for these molecules, since this parameter determines the reemission probability after absorption of scintillation light in the liquid. The absorption of secondary wavelength shifters like bis-MSB becomes negligible in the region above 430~nm. There the impurity levels in the chemicals are most relevant for the attenuation length. Light absorbed by the impurities is typically not reemitted. The regions of main absorption for the individual LS components are illustrated in Figure~\ref{Fig1}.  

\begin{figure}[tb]
\begin{center}
\includegraphics[scale=0.50]{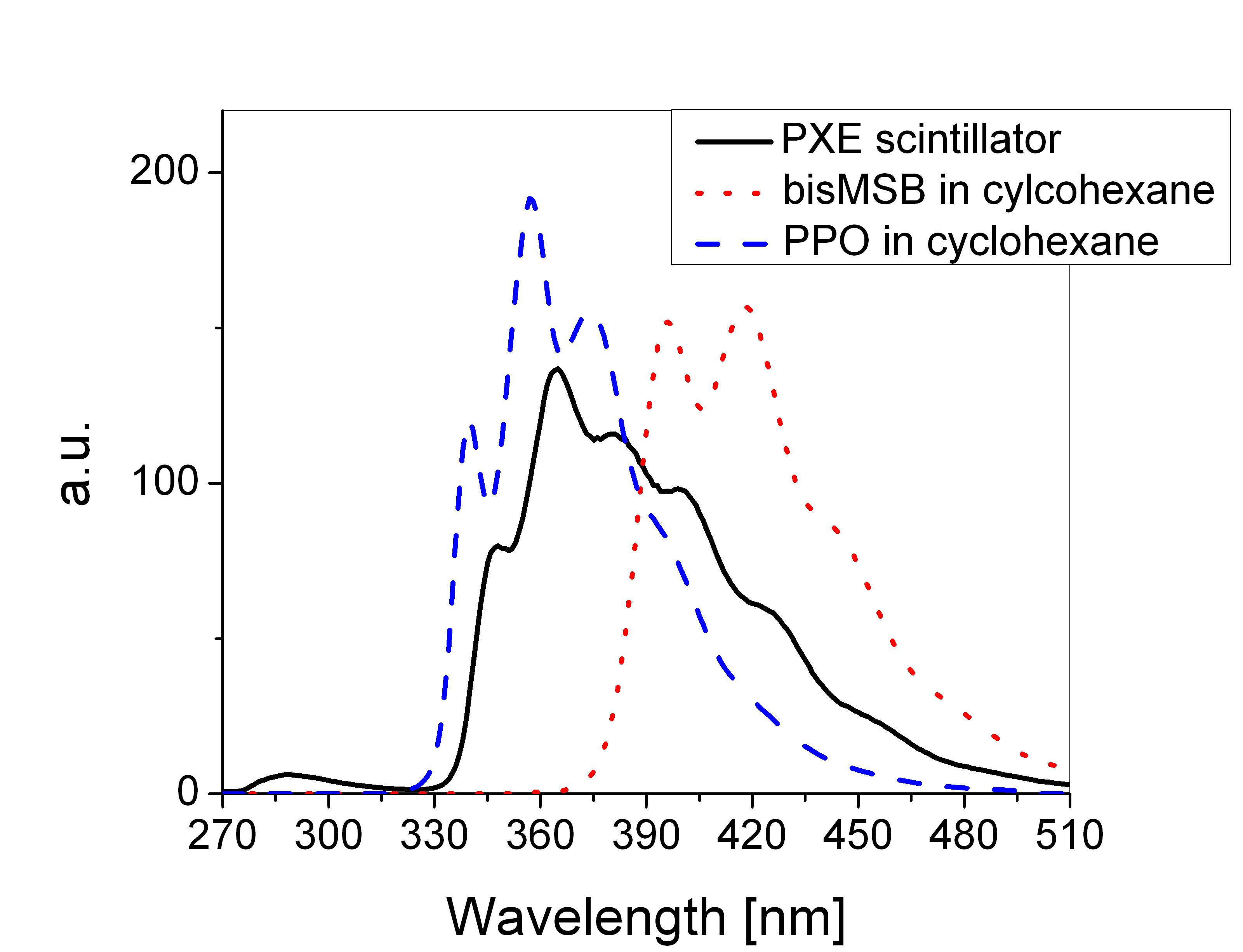}
\end{center}
\caption[]{Emission spectra of PPO and bis-MSB samples diluted in cyclohexane are compared to a o-PXE based scintillator (3~g/l PPO, 20~mg/l bis-MSB) spectrum measured in a geometry with negligible self absorption.}   
\label{Fig2}
\end{figure}

The emission spectra of PPO and bis-MSB diluted in cyclohexane at concentrations of 0.5~mg/l are shown in Figure~\ref{Fig2}. In addition, the emission spectrum for a LS mixture as measured in a 2~mm cell in the so-called front face geometry is plotted. For this geometry, emission spectrum of the sample is recorded at the same side where the excitation takes place. The sample contains 3~g/l PPO and 20~mg/l bis-MSB dissolved in pure ortho-phenylxylylethane (o-PXE), a high flashpoint scintillator solvent used e.g.~in the Borexino CTF~\cite{PXE} and in the Double Chooz experiment~\cite{DC1}. This sample was excited at 260~nm, where the light mainly excites the o-PXE. 

The LS emission spectrum in Figure~\ref{Fig2} shows the basic features of a pure PPO spectrum shifted by 8~nm to longer wavelengths. This wavelength shift can be explained by the combination of a solvent and concentration effect. In the spectrum, there is also some contribution of the bis-MSB (less than 30\%) and a small PXE emission peak around 290~nm. We mainly see the PPO emission when exciting the PXE due to the fact that there is non-radiative energy transfer from the excited PXE to the PPO whereas the energy transfer from PPO to bis-MSB is mainly radiative. Since propagation through the liquid is minimized for the detected light in the front face geometry, radiative transfer is suppressed in this case. The rather poor non-radiative transfer between the two wavelength shifters due to a slight mismatch in the main PPO emssion and bis-MSB absorption could be improved by using a different primary wavelength shifter like e.g.~p-terphenyl~\cite{PXE} which on the other hand has the disadvantage of limited solubility. 

In detector simulations it is normally better to use the emission spectrum after non-radiative transfer instead of the ones of the single molecules~\cite{SWD}. Otherwise the simulation might assume radiative energy transfer, which differs from non-radiative transfer in several aspects as e.g.~the efficiency. Non-radiative transfer is much harder to simulate due to contribution of multiple processes and complicated microphysics~\cite{Ent}. 

\begin{figure}[tb]
\begin{center}
\includegraphics[scale=0.50]{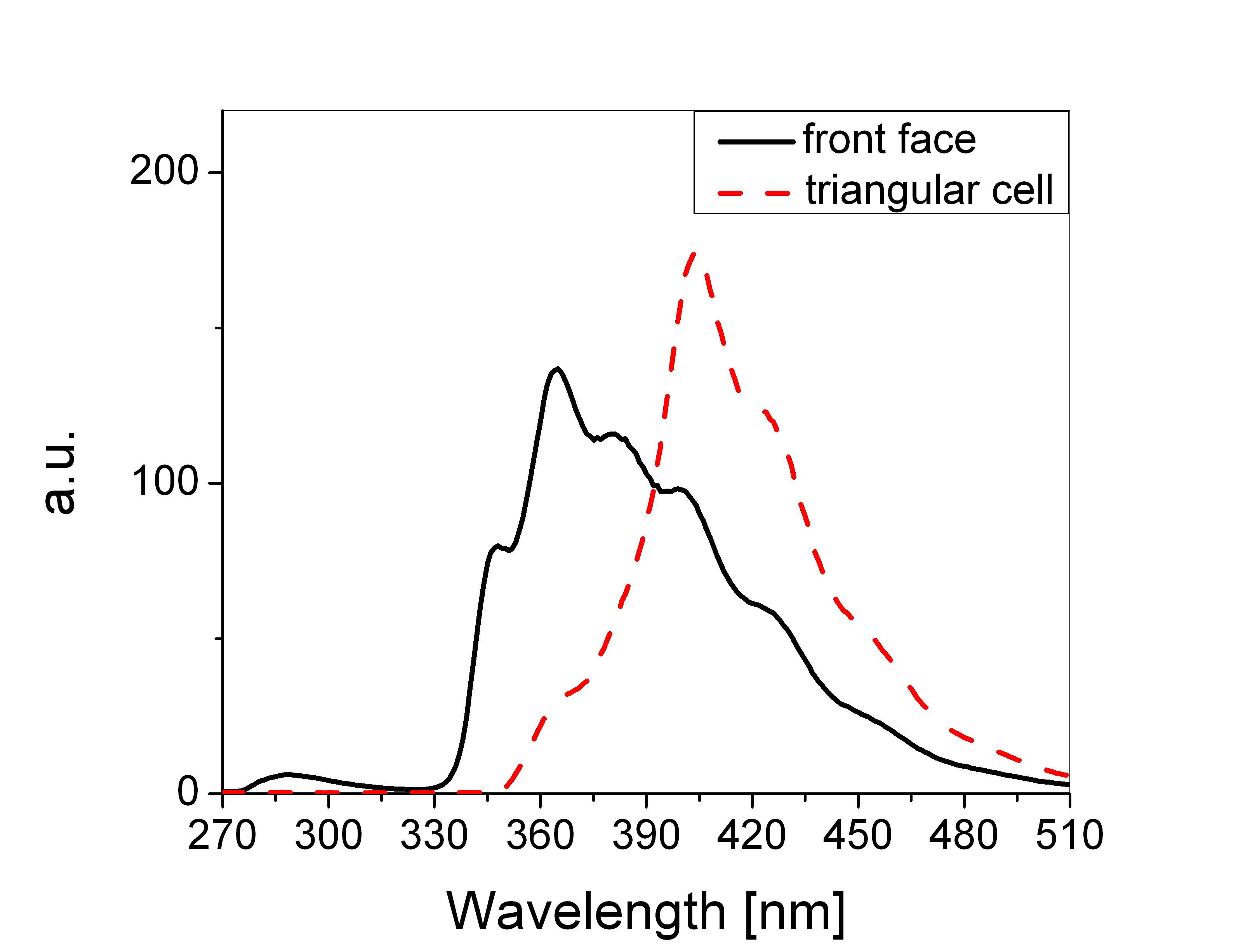}
\end{center}
\caption[]{Scintillator emission in front face geometry (small self-absorption) and in a 1~cm triangular cell (some self-absorption).}  
\label{Fig3}
\end{figure}

As the light propagates through the liquid and gets absorbed and reemitted by the secondary wavelength shifter, the emission spectrum changes and shifts to longer wavelengths, since the absorbed photons can only be reemitted at equal or lower energies. Figure~\ref{Fig3} shows again the LS spectrum of Figure~\ref{Fig2}. In addition, the spectrum of the same PXE based scintillator mixture is plotted for a case when the light had to pass through the liquid for few millimetres. In this spectrum the PPO component has essentially disappeared and the light is mainly emitted in the bis-MSB region above 400~nm. At longer pathlengths of several centimetres or even metres the spectrum will be even more shifted to longer wavelengths. To predict the response of light sensors such as photomultiplier tubes (PMTs) the wavelength dependent quantum efficiency of the device as well as the light spectrum at the PMT position needs to be known. As we see from Figure~\ref{Fig3}, the distance between the point of light creation and PMT position is relevant for the LS emission spectrum. 

Scintillation photons from $350-430$~nm are mainly absorbed by the secondary wavelength shifter as described above. At shorter wavelength the amount of scintillation light is negligible. Nevertheless, the reemission probability below 350~nm is of interest as well, in case of Cherenkov light production. Good modelling of the Cherenkov contribution can be of crucial importance to understand the non-linearity in the energy scale of a detector~\cite{SWD}. Photons emitted after light absorption below 350~nm mainly originate from excited PPO molecules. Here the PPO molecules are excited either directly ($280 - 350~$nm) or indirectly via non-radiative energy transfer from the solvent ($< 280$~nm). Nevertheless, the reemission probability does not only depend on the fluorescence quantum yield of PPO, in particular for wavelengths lower than 300~nm. It is also influenced by the interaction mechanisms between molecules. The reemission probability in the LS has a rising trend with increasing wavelength and converges to the fluorescence quantum yield of PPO before bis-MSB absorption and emission take over~\cite{SWD}. Therefore the quantum yield not only of the secondary, but also of the primary wavelength shifter is an important parameter for the modelling of Cherenkov light. The secondary wavelength shifter is added typically at low concentrations to the scintillator and the emission spectrum is shifted relative to the absorption bands of the other molecules. Therefore, the reemission probability above 350~nm is less affected by self-interactions or solvent effects and can be approximated by the bis-MSB fluorescence quantum yield.
     
\section{Fluorescence quantum yields}

The literature values of fluorescence quantum yields, in particular for PPO and bis-MSB, vary over a wide range. In many databases the values of Berlman~\cite{Ber} are quoted. However, these values might overestimate the true numbers. For the quantum yield of one of the samples ``Berlman has rather arbitrarily assigned a value of 1.00''~\cite{Dem}. This sample was used as a standard and all other numbers in the Berlman~\cite{Ber} handbook refer to it. The literature values we found for PPO range from 0.71 to 1.00, those for bis-MSB from 0.75 to 0.96 \cite{Ber, XIA, Mas, Joh}.

The reasons why it is so hard to reproduce those results are manifold. The determination of the quantum yield strongly depends on many parameters such as sample concentration, solvent choice, oxygen content, excitation/emission wavelength dependencies, sample geometry, self-absorption or instrumental effects. Since the values are typically determined in relative measurements, a rather large uncertainty on the knowledge of the quantum yield standard also enters the measurement. Therefore it is useful to have several independent measurements using different setups. 

For our measurements we diluted the samples in cyclohexane. This rather inert solvent has the advantage of high transparency and optical purity. The PPO and bis-MSB samples were diluted to a concentration of 0.5~mg/l. The absorption spectrum was measured using a Varian Cary400 UV/Vis photospectrometer. To get the emission spectrum a Varian CaryEclipse fluorimeter was used. The measurements were done in a 1~cm quartz cell. Oxygen was removed before the measurements by nitrogen bubbling. After bubbling the cells were closed with a sealed cap. To avoid any influence due to variation in the light intensity at different excitation wavelengths we extracted the quantum yield using the same excitation wavelength in the sample and the standard. As a standard we were using quinine sulfate ($1.5\cdot 10^{-5}$~M) in 0.1~M sulfuric acid. 

This quinine sulfate reference fulfills several criteria of a adequate standard as good separation of broad absorption and emission bands, which are in similar wavelength regions as those of the samples. Furthermore, it is rather insensitive to oxygen or concentration quenching and stable in solution. Its quantum yield $\Phi_r$ of 0.55~\cite{Mel} is rather accurately known and constant over the investiagted excitation wavelengths (290 -- 380~nm)~\cite{Gil}. However, there is a temperature dependence of the yield~\cite{Mel}. The measurements were done at a room temperature of about $21 - 25^\circ$C. In this range corrections to the reported quantum yield of the standard should be less than 1~\%.      

Since the solvent in the sample and the solvent in the standard are different, a refractive index correction had to be applied. The quantum yields of the samples $\Phi_x$were calculated using 
\begin{equation}
\Phi_x = \Phi_r \frac{B_r}{B_x} \frac{I_x}{I_r} \frac{n_x^2}{n_r^2}.
\end{equation}
In this equation $B_x$ and $B_r$ correspond to the fraction of light absorbed in the sample ($x$) and the standard ($r$). The values for $B$ were kept below $20$~\% in sample and reference. The  wavelength integrated intensities of the emission spectra $I_x$ and $I_r$ were corrected for the instrumental wavelength dependent detection efficiency. For the refractive index of cyclohexane we used $n_x = 1.43$ and for the one of the standard $n_r = 1.34$.  

The PPO quantum yield was determined at an excitation wavelength of 290~nm and from 300 to 330 nm in 5~nm steps. An average value of 0.842 was found. For bis-MSB the result we obtain is 0.863 which is also an average in a wavelength range from 300 to 380 nm (5~nm steps from 300 -- 350~nm, 10~nm steps from 350 -- 380~nm). Both values are found to be stable within the wavelength range covered by the measurements. The variations from the average value was within 3~\% for all wavelengths. As an additional data point the PPO yield was checked at an excitation wavelength of 350~nm. To obtain a reasonable absorption at this wavelength the PPO-concentration in cyclohexane was increased by more than 2 orders of magnitude to $4.5\cdot 10^{-4}$~M. Even at these high concentrations no indication for any decrease in the quantum yield could be found in the tail of the absorption spectrum.  

Since the emission of the bis-MSB light is mainly above 400~nm it would be interesting to test the yield also in that region to get the reemission probability when the photons are self-absorbed. However, the bis-MSB absorption drops rapidly in that area and much higher concentrations are needed to determine those values. This might induce systematic effects influencing the precision of the results. In addition, the quantum yield of the standard is not reliable anymore above 400~nm and one would need to compare it to data taken at lower excitation wavelengths, which requires additional correction factors increasing the uncertainty further. For those reasons we restricted ourselves in the determination of the quantum yields to wavelengths up to 380~nm. 

To check systematic effects the measurements were repeated varying some of the relevant parameters. To investigate effects of concentration quenching or self-absorption the yield was determined at lower concentrations (factor 2). In addition, the cell geometry was modified using also a thinner cell of 4~mm length and 1~cm width. The results obtained all agree within a $1 \sigma$ error of 5~\%, which was estimated from the systematic contributions as listed above. In particular, this 5~\% relative error covers the uncertainties related to the precision of the reference yield, knowledge of the refractive indices and quenching effects due to inefficient oxygen removal. After oxygen removal in our samples the quantum yield increased in the case of bis-MSB by 7~\%. A weaker effect was found for PPO.  

The ratio of the bis-MSB to PPO quantum yield is 1.02 for our results. This is in reasonable agreement to the ratio of Berlman which is 0.94~\cite{Ber} assuming the Berlman ratio to also have an uncertainty in the 5~\% range. However, our absolute values are significantly lower, which might be explained by the fact that Berlman overestimated the yield of his reference, as already suggested by Demas and Crosby~\cite{Dem}. On the other hand our results are higher than the ones measured by members of the Borexino Collaboration~\cite{Joh, Mas}, but also here we have reasonable relative agreement for the $\Phi_\textrm{\scriptsize bis-MSB}/\Phi_\textrm{\scriptsize PPO}$ ratio. If we compare the results presented here with the values of Xiao et al.~\cite{XIA} we find good agreement for the absolute value of PPO, however, the yield for bis-MSB is about 10~\% lower in our case. Part of the reason could be a strong solvent effect, since the solvent for bis-MSB in reference~\cite{XIA} is linear alkyl benzene (LAB) instead of cyclohexane. 

\vspace{0.2 cm}

To study such solvent effects we also did the measurements replacing cyclohexane by \linebreak \mbox{n-dodecane}, LAB and o-PXE. The values in n-dodecane (the refractive index $n_\textrm{\scriptsize dodecane}$ at 405~nm, 18$^\circ$C is 1.42), the main component in the KamLAND~\cite{Kam} and Double Chooz detectors~\cite{LSP}, were within the precision of the measurement consistent with the ones determined in cyclohexane. The aromatic solvent LAB ($n_\textrm{\scriptsize LAB} = 1.49$) which is e.g.~used in the targets of the Daya Bay~\cite{DB} and RENO~\cite{RENO} reactor neutrino experiments has a rather strong self emission when excited below 340~nm making the measurement complicated. Since the PPO absorption is already very weak above 340~nm we only determined the yield for bis-MSB in LAB from 340 to 380~nm. At a bis-MSB concentration of about $1\cdot 10^{-6}$~M the yields were 5 to 10~\% higher in LAB compared to cyclohexane which is consistent with the bis-MSB result reported by Xiao et al.~\cite{XIA}. For o-PXE ($n_\textrm{\scriptsize PXE}$ at 405~nm, 18$^\circ$C is 1.604) which is used in Double Chooz there are similar difficulties at lower wavelengths as for the LAB, so also here the measurement was only done for bis-MSB at 370 and 380~nm. Slightly lower yields were recorded in o-PXE compared to cyclohexane, but the effect was only 5~\% or less and therefore not significant. In reference~\cite{Joh} quite large solvent effects were reported in pseudocumene (1,2,4-trimethylbenzene, PC) which is the basis of the Borexino~\cite{Bor} scintillator and also contained in the KamLAND detector. Compared to the cyclohexane solvent the value increased by 13~\% for the case of PPO and it decreased by 13~\% for bis-MSB.

\vspace{0.2 cm}

In Table~\ref{QY_sum} we summarize all our results obtained in cyclohexane for several solvent or wavelength shifting molecules and other fluorescence materials of interest. Anthracene and 9,10-di-phenylathracene (DPA) are sometimes used as standards in this type of measurements. Both of them have the disadvantage that they are very sensitive to oxygen quenching and that there absorption bands are very narrow and spiky making them sensitive to instrumental parameters as the excitation bandwidth. Besides, the separation of absorption and emission bands is rather poor increasing self-absorption processes inside the sample cell. So we observed in the DPA sample a rather high spread in the quantum yield numbers for the different excitation wavelengths and found a relative difference in the integrated light emission with and without nitrogen bubbling of almost 20~\%. 

In many studies anthracene is diluted in ethanol, so we determined the yield also in this solvent ($n = 1.36$). Values were measured between 300 and 370~nm in 5~nm steps and the yields were found to be stable within errors in this region. For anthracene in ethanol ($1.1\cdot 10^{-5}$~M) $\Phi = 0.297\pm0.022$ was found which is in agreement to the literature value of 0.27~\cite{Dem}. Dissolved in cyclohexane ($1.1\cdot 10^{-5}$~M) we get a higher value of $\Phi = 0.323\pm0.021$. This number can be compared to the one in the handbook of Berlman which is 0.36. Our lower result is consistent with the suspicion that all of Berlman's values should be reduced.       

\begin{table}[h]
\caption[Composition]{Fluorescence quantum yield of aromatic molecules used in liquid scintillators. All samples were diluted in cyclohexane and measured at room temperature inside a 1~cm fluorescence cell (for DPA, BPO, PBD and butly-PBD the yields determined in the 4~mm cell were used). Besides the range of excitation wavelengths used to determine the quantum yields the wavelengths of the absorption and emission peaks are quoted.} \label{QY_sum}
\begin{center}
\begin{tabular}{lccccc}
Molecule & concentration & exc.~range~[nm] & Quantum yield & abs.~max. & em.~max.\\
\hline
bis-MSB &  $1.6\cdot 10^{-6}$~M & 300 - 380 & $0.863\pm0.043$ & 345~nm & 418~nm \\
PPO	&  $2.3\cdot 10^{-6}$~M &  290 - 330 & $0.842\pm0.042$ & 303~nm & 358~nm \\
Anthracene &  $1.1\cdot 10^{-5}$~M &  300 - 370 & $0.323\pm0.021$ & 357~nm & 400~nm\\ 
DPA & $1.5\cdot 10^{-6}$~M & 350 - 380 & $0.91\pm0.05$ & 373~nm & 407~nm \\ 
BPO & $1.7\cdot 10^{-6}$~M & 320 - 350 & $0.91\pm0.05$ & 320~nm & 384~nm \\
PBD & $1.7\cdot 10^{-6}$~M & 300 - 320 &  $0.84\pm0.05$  & 302~nm & 358~nm \\
butyl-PBD & $1.4\cdot 10^{-6}$~M & 300 - 320 &  $0.89\pm0.05$ & 302~nm & 361~nm \\
POPOP &  $1.4\cdot 10^{-6}$~M  & 350 - 380 & $0.90\pm0.05$ & 360~nm & 411~nm \\
o-PXE & $1.2\cdot 10^{-4}$~M & 260 - 280 & $0.33\pm0.03$ & 269~nm & 290~nm \\
PC & $8.3\cdot 10^{-5}$~M & 250 - 270 & $0.41\pm0.04$ & 267~nm & 290~nm \\
LAB & $4.1\cdot 10^{-4}$~M & 250 - 270 & $0.20\pm0.02$ & 260~nm & 284~nm \\
DIN & $4.7\cdot 10^{-5}$~M & 250 - 270 & $0.32\pm0.03$ & 279~nm & 338~nm \\ 
\hline
\end{tabular}
\end{center}
\end{table}

Among all the molecules studied by Berlman as well as in our studies the highest fluorescence quantum yield is found for DPA. This yield was set to 1.00 in the Berlman book and used as reference for all other molecules. At excitation around the main absorption peak (350~nm to 380~nm), we find a quantum yield of 0.91 measured in a 4~mm fluorescence cell. This means if compared to the results presented here all of Berlman's values should be reduced by 9~\%.
 
Alternative options for PPO or bis-MSB in organic LS were measured for comparison. The primary wavelength shifters BPO (2-(4-biphenyl)-5-phenyloxazole), PBD (2-(4-biphenyl)-5-phenyl-1,3,4-oxadiazole) and butyl-PBD (2-(4-biphenyl)-5-(4-tert-butyl-phenyl)-1,3,4-oxadiazole) were already investigated in the context of Indium loaded LS \cite{JL}. For PBD and butyl-PBD the reference sample was the diluted PPO-solution which absorbs and emits in very similar wavelength regions. For the primary wavelength shifter candidates the highest values are found for BPO and butyl-PBD. Liquid scintillators containing these molecules typically also provide high light yields~\cite{JL}. Although there is reasonable relative agreement between quantum yield ratios for most molecules investigated in this work with the available literature values of Berlman~\cite{Ber} and most discrepancies in the absolute values could be explained by a common correction factor of about 10~\%, there is a notable difference between the PBD and the PPO quantum yield. We find identical numbers whereas Berlman finds a PBD quantum yield which is 17~\% below the one of PPO. 

A possible replacement of the secondary wavelength shifter bis-MSB is POPOP (5-Phenyl-2-[4-(5-phenyl-1,3-oxazol-2-yl)phenyl]-1,3-oxazole) which seems to have a slightly higher yield.   

For the solvent molecules in Table~\ref{QY_sum} the estimated relative uncertainty is higher than the one of the wavelength shifters, since absorption and emission bands are shifted to lower wavelengths compared to the standard. In addition, those solvent molecules are very sensitive to oxygen quenching. Whereas several neutrino experiments in the past used PC-based scintillators providing high light yields, more recent experiments prefer high flash-point solvents which are advantageous in terms of safety aspects as o-PXE or LAB. The yield of o-PXE (Dixie Chemicals) was first determined relative to the quinine sulfate reference. The other solvents were then measured relative to PXE. For PC (Aldrich, 98~\%) the highest yield of all tested solvents was found. The LAB (Petresa 550-Q) had significantly lower yields than the other molecules. Another solvent with properties which are promising for large scale neutrino detectors, Diisopropyl naphtalene (DIN, Ruetasolv DI-S, mixture of isomers), has a similar yield as o-PXE.   

\section{Conclusion}

For a good modelling of the light propagation in large scale liquid scintillator detectors a detailed knowledge of the light absorption properties, the photon emission behaviour and the fluorescence quantum yields of the wavelength shifters is essential. For the commonly used molecules PPO and bis-MSB the literature values for the quantum yield vary by 30~\%. Therefore new measurements were required with an emphasis on the control over systematic effects as presented in this article. The quantum yields obtained in our studies are $\Phi_\textrm{\scriptsize PPO} = 0.842\pm0.042$ and $\Phi_\textrm{\scriptsize bis-MSB} = 0.863\pm0.043$. Similar values ranging from $84-91$~\% were found in alternative primary and secondary wavelength shifters when dissolved in cyclohexane. Variations of the scintillator solvent can change the yield by about 10~\% in some cases.


\begin{thebibliography}{99}

\bibitem{Ber} I.B.~Berlman, \emph{Handbook of fluorescence spectra of aromatic molecules}, Academic Press, New York and London (1971).

\bibitem{PXE} H.O.~Back et al., \emph{Study of phenylxylylethane (PXE) as scintillator for low energy neutrino experiments}, Nucl.~Instrum.~Meth.~A 585 (2008) 48 -- 60.

\bibitem{DC1} Double Chooz Collaboration, \emph{Indication for the disappearance of reactor $\bar{\nu}_e$ in the Double Chooz experiment}, Phys.~Rev.~Lett.~108 (2012) 131801.

\bibitem{SWD} S.~Wagner, \emph{Energy non-linearity studies and pulse shape analysis of liquid scintillator signals in the Double Chooz experiment}, PhD Thesis, Universit\"at Heidelberg 2014.

\bibitem{Ent} C.~Aberle, C.~Buck, F.X.~Hartmann, S.~Sch\"onert, \emph{Light yield and energy transfer in a new Gd-loaded liquid scintillator}, Chemical Physics Letters 516 (2011) 257 -- 262.

\bibitem{Dem} J.N.~Demas and G.A.~Crosby, \emph{The Measurement of Photoluminescence Quantum Yields. A Review.}, The Journal of Physical Chemistry, Vol.~75, 8 (1971) 991 -- 1024.

\bibitem{XIA} H.~Xiao, X.~Li, D.~Zheng, J.~Cao, L.~Wen, N.~Wang, \emph{Study of absorption and re-emission processes in a ternary liquid scintillation system.}, Chinese Physics C, Vol.~34, 11 (2010) 1724 -- 1728.

\bibitem{Mas} F. Masetti, F. Elisei, U. Mazzucato, \emph{Optical study of a large-scale liquid-scintillator detector}, Journal of Luminescence 68 (1996) 15 -- 25. 

\bibitem{Joh} M.C.~Johnson, \emph{Scintillator Purification and Study of Light Propagation in a Large Liquid Scintillator Detector}, PhD thesis, Princeton University (1998).

\bibitem{Mel} W.H.~Melhuish, \emph{Quantum efficiencies of fluorescence of organic substances: effect of solvent and concentration of the fluorescent solute}, J.~Phys.~Chem.~65 (1961) 229.

\bibitem{Gil} J.E.~Gill, \emph{The fluorescence excitation spectrum of quinine bisulfate}, Photochem.~Photobiol.~9, 313 (1969).

\bibitem{Kam} KamLAND Collaboration, \emph{First Results from KamLAND: Evidence for Reactor Antineutrino Disappearance}, Phys.~Rev.~Lett.~90, 2 (2003) 021802.

\bibitem{LSP} C.~Aberle et al., \emph{Large scale Gd-beta-diketonate based organic liquid scintillator production for antineutrino detection}, JINST 7 (2012) P06008.

\bibitem{DB} Daya Bay Collaboration, \emph{Observation of Electron-Antineutrino Disappearance at Daya Bay}, Phys.~Rev.~Lett.~108 (2012) 171803.

\bibitem{RENO} RENO Collaboration, \emph{Observation of Reactor Electron Antineutrinos Disappearance in the RENO Experiment}, Phys.~Rev.~Lett.~108 (2012) 191802.

\bibitem{Bor} Borexino Collaboration, \emph{Direct Measurement of the 7Be Solar Neutrino Flux with 192 Days of Borexino Data}, Phys.~Rev.~Lett.~101 (2008) 091302.

\bibitem{JL} C.~Buck et al., \emph{Luminescent properties of a new In-based organic liquid scintillation system}, Journal of Luminescence 106 (2004) 57 -- 67.

\end{thebibliography}
\end{document}